# A SECURE KEY TRANSFER PROTOCOL FOR GROUP COMMUNICATION

R. Velumadhava Rao[1], K Selvamani[2], R Elakkiya[3]

[1]Department of Computer Science, Rajalakshmi Institute of Technology, Chennai, India
Velu_b4u@yahoo.com
[2]Department of Computer Science, Anna University, CEG, Chennai, India
smani@cs.annauniv.edu
[3]Department of Computer Science, Jerusalem Engineering College, Chennai, India
elakkiyaceg@gmail.com

## ABSTRACT

*Providing security for messages in group communication is more essential and critical nowadays. In group oriented applications such as Video conferencing and entertainment applications, it is necessary to secure the confidential data in such a way that intruders are not able to modify or transmit the data. Key transfer protocols fully rely on trusted Key Generation Center (KGC) to compute group key and to transport the group keys to all communication parties in a secured and secret manner. In this paper, an efficient key generation and key transfer protocol has been proposed where KGC can broadcast group key information to all group members in a secure way. Hence, only authorized group members will be able to retrieve the secret key and unauthorized members cannot retrieve the secret key. Hence, inorder to maintain the forward and backward secrecy, the group keys are updated whenever a new member joins or leaves the communication group. The proposed algorithm is more efficient and relies on NP class. In addition, the keys are distributed to the group users in a safe and secure way. Moreover, the key generated is also very strong since it uses cryptographic techniques which provide efficient computation.*

## KEYWORDS

*Key Distribution, Forward Secrecy; Backward Secrecy; Message Confidentiality*

## 1. INTRODUCTION

Message Confidentiality is one of the most important features in secure group communication. Message confidentiality ensures that the sender confidential data which can be read only by an authorized and intended receiver. Hence, the confidential data is secured in efficient way such that it is not tampered by unauthorized users. Message Confidentiality is achieved mainly by two components namely Forward Secrecy and Backward Secrecy are used for message confidentiality. The two main components provide secrecy as follows.

- *Forward Secrecy:* Forward Secrecy is a mechanism to ensure that whenever the user leaves the group he will not have any access to the future key.
- *Backward Secrecy:* Backward Secrecy ensures that whenever a new user joins the group, he will not get any access to the previous details.

The most common method used for achieving group communication in a secure way is by encryption techniques. To provide a secure group communication, it is necessary to manage the keys in a secure way for creating, updating and distribution of those keys. Moreover, before exchanging the confidential data, the key establishment protocol has to distribute the group key to all group entities in a secured and effective manner. The two important types of key establishment protocols used to achieve this are namely key transfer protocols and key agreement protocols. Key transfer protocols rely on KGC to select group key for communicating information with the group members by sharing one or more secret key during registration. But in key agreement protocols, the group key is determined by exchanging public keys of two communication parties with the presence of communication entities.

The most commonly used key agreement protocol in the existing work is Diffie-Hellman (DH) key agreement protocol [2]. However, the Diffie Hellman key distribution algorithm is limited to provide secret key only for two entities, and cannot provide secret keys for the group that has more than two members. When there are a more number of members in a group the time delay for setting up the group key will take longer time. Hence, it is necessary to propose a new technique to avoid this type of constraints in group communication. In this proposed work, group communication applications will make use of key transfer protocol to transmit data to all the group members with the minimum resources needed for this group communication. In the existing approach, the distributed group key management protocol does not have the group controller that controls the users in the group for secure group communication. Also, the processing time and communication requirements increase when the number of users increases in the group. Hence, it is necessary to propose a new technique that ensures the group communication among the group using forward secrecy and backward secrecy, while avoiding the existing overheads. Moreover, the members joining and leaving the group are efficiently managed by encrypted key. Since, the key is generated and maintained for each member in a group the confidentiality and security of the data in group communication is guaranteed. Also, the computation time of the message distribution and key extraction from the message is fast, when compared with the existing work.

In this work, it uses cryptographic techniques to secure key distribution and key management for the group communication environment. The algorithms are analyzed with suitable samples. The remainder of this paper is organized as follows. Chapter 2 surveys about the existing work in this area. Chapter 3 explores the proposed work and Chapter 4 explores about the implementation work. Chapter 5 analyzes and discusses the results obtained from the work. Chapter 6 concludes the proposed and implemented work and suggested some possible enhancements. .

## 2. RELATED WORK

There are many works pertaining to the secure Group Communication and that have been carried out, but some of the important works has been surveyed and cited here. Among them, Mike Burmester and Yvo Desmedt [3] presented a Group Key Exchange protocol which extends the Diffie-Hellman protocol [2]. The protocol is scalable and secure against passive attacks. But, Diffie Hellman public key distribution algorithm is able to provide group key only for two entities. Bohli [4] developed a framework for robust group key agreement that provides security against malicious insiders and active adversaries in an unauthenticated point-to-point network. Bresson et al. [13] constructed a generic authenticated group Diffie-Hellman key exchange algorithm which is more secure. Katz and Yung [5] proposed the first constant-round and fully scalable group Diffie-Hellman protocol which is provably secure. There are many other works related to group key management protocols based on non-DH key agreement approaches. Among them, Tzeng [9] presented a conference key agreement protocol that relies on discrete algorithm assumption with fault tolerance. This protocol establishes a conference key even if there is several numbers of malicious participants in the conference. Hence, this method is not suitable for group communication.

Moreover, in a centralized group key management, there is only one trusted entity responsible for managing the entire group. Hence, the group controller need not depend on any auxiliary entity to perform key distribution. Harney et al. [10] proposed a group key management protocol that requires O(n) where n is the size of group, for encrypting and update a group key when a user is evicted or added in backward and forward secrecy.

Eltoweissy et al.[6] developed a protocol based on Exclusion Basis Systems (EBS), a combinatory formulation for the group key management problem. Lein Harn and Changlu Lin [1] introduced a group key transfer protocol where members of the group fully rely on Key Generation Center (KGC). They proposed an authenticated key transfer protocol based on secret sharing scheme that KGC can broadcast group key information to all group members at once. Chin-Yin Lee et al. [7] addressed the security issues and drawback associated with existing group key establishment protocols. They have also used secret sharing scheme to propose a secure key transfer protocol to exclude impersonators from accessing the group communication. Their protocol can resist potential attack and also reduce the overhead of system implementation. Burmester et.al [11] has presented a practical conference key distribution systems based on public-keys and also authenticates the users. Rafael Martinez Pelaez et.al [14] proposed a new dynamic ID-based remote user authentication scheme based on Hsian-Shih's scheme. In the proposed scheme users can create the login request message without known the identification of each server and also it is more efficient in computation cost.

## 3. PROPOSED WORK

Based on the literature survey, it is necessary to propose a new model that communicates data in a safe and secured manner in group communication. The proposed model consists of four processes namely the User Registration, Group key generator based on prime numbers, Key generation and Key distribution, Group re-keying. The four main processes are explained as below.

### 3.1. User Registration

This module explains the process of User Registration and Key Computation. Each user has to register their identity at KGC for subscribing the key distribution service. KGC keeps track of all registered users and removes any unsubscribed users in the group. During registration process, each user Ui is required to share a random secret value Si with the KGC. Once user registration process is completed, KGC assigns a permanent secret id, denoted by Pi for each member Ui in the group.

### 3.2. Group Key Generator Based on Prime Numbers

In group key generator, whenever there is a group of users participating in a group communication, the Key Generation Center (KGC) will generate the message M. When there are N members in the group such as N= ($U_1$, $U_2$,….., $U_n$), then each member $U_i$ will be assigned a $P_i$, a large prime number > K where K is a group key.

### 3.3. Key Generation and Distribution

When a request is made by the user for group key generation, KGC randomly selects a group key K whose value is less than Pi where (i = 1, 2 ..., n). KGC needs to distribute the group key in a secure manner. Suppose if the group consist of n members, {$U_1$, $U_2$,..., $U_n$} and the shared primes are {P1, P2,.., Pn}. The group key can be any prime number or any number K < $P_i$ $\forall i$. That is $P_1$>K, $P_2$>K,..., $P_n$>K.

KGC now generate a message M and it's value is calculated using a formula

$$M = ((P_1 \oplus S_1)*(P_2 \oplus S_2),...,*(P_n \oplus S_n)) + K \qquad (1)$$

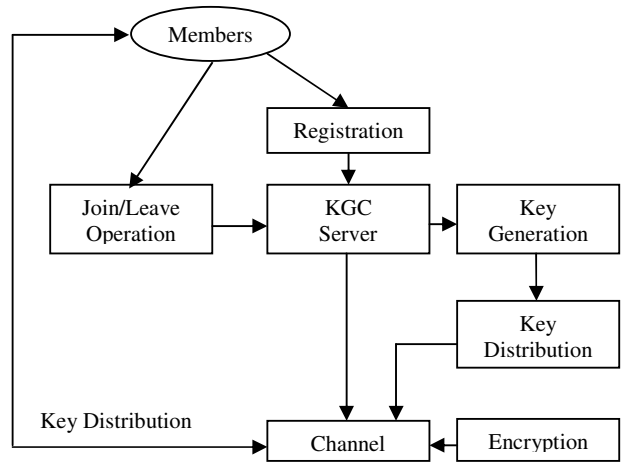

Figure 1.  System Architecture

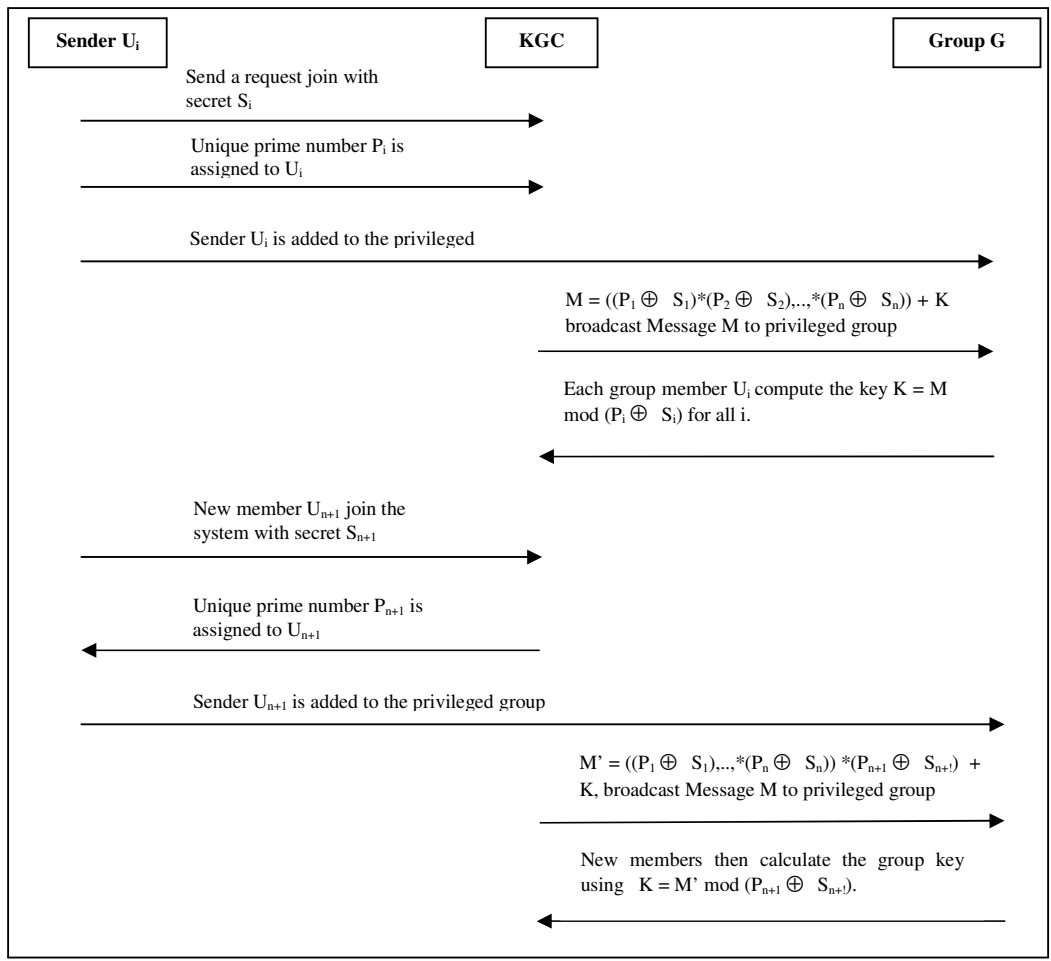

Figure 2. Process of secure key transfer

The Prime number Pi is XORed with Secret value Si where i= 1,2,...,n. Each XORed values are multiplied and added along with the Key K. Now message M is broadcasted to all the members in a group. Upon receiving the message M, the each member in the groups will generate the key in the following manner.

$$K = M \bmod (Pi \oplus Si) \text{ for all i.} \qquad (2)$$

Each member in the group generates the Key by doing XOR operation with prime number and the secret value. Then the modulus will be taken for the message M along with the value calculated from the XORed operation.

## 3.4. Group Re-keying

Scalable group re-keying is the important task to be performed when user joins or leaves the group in the secure group communication. The group keys needs to be updated to maintain the forward and backward secrecy. To achieve this, the two important tasks namely members join and members leave operation is performed.

### 3.4.1. Member Join

When a member wants to join the group, the new member will register with the KGC. KGC will share a prime number $P_n+1$ where $K < P_n+1$. KGC generates the new key using the formula.

$$M' = ((P_1 \oplus S_1)*(P_2 \oplus S_2),..,*(P_n \oplus S_n)) *(P_n+1 \oplus S_n+1) + K \qquad (3)$$

The prime number Pi and the secret value Si where i=1,2,...n is XORed. Each XORed values are multiplied and then the group Key K is added along with that.Then the new member calculates the group key by using the message M'.

$$K = M' \bmod (P_n+1 \oplus S_n+1) \qquad (4)$$

The new prime number $P_n+1$ and secret value $S_n+1$ is XORed and modulus will be taken along with the new message M'.

### 3.4.2. Member Leave

When a member leaves the system, the member should inform to the KGC. Now KGC generates a new group key as follows.
Step 1. KGC selects a new prime number K' (where $K' < P_i$ for all i).
Step 2. New message M'' is calculates using a formula.

$$M'' = ((P_1 \oplus S_1)*(P_2 \oplus S_2),..,*(P_{j-1} \oplus S_{j-1})) ((P_{j+1} \oplus S_{j+1})*... (P_n \oplus S_n)) + K' \qquad (5)$$

The values of Pi and Si where i=1,2,..,n are XORed and the XORed values are multiplied also added along with the new group Key K'. Now KGC broadcast a new message M'' to all group members. Upon receiving the message the group key is calculated by each member in the following manner

$$K = M'' \bmod (P_i \oplus S_i) \qquad (6)$$

In addition to this, the process for secure key transfer in group communication is shown in Figure 2. The members who wish to join the group send a request with secret identity $S_1$ which is responded by KGC with unique prime number $P_1$. KGC generates message M using the formula (1) and broadcast M to the privileged group members. Each group member is able to compute the group key K using the formula (2). Similarly, whenever member joins/leaves the system group key K and message M is regenerated.

## 4. EXPERIMENTAL SETUP

The experimental setup consists of group of 10 members. Members 1,2,..,10 has the permanent prime id (Prime) as, P1=55837, P2=55603, P3=35353, P4=54709, P5=60799, P6=45953, P7=40847, P8=39461, P9=42709, P10=58909 respectively along with the group key K= 22971. Moreover, the secret key shared by the users are S1=28931, S2=37123, S3=12347, S4=13745, S5=16231, S6=31234, S7=21467, S8=25431, S9=17237, S10=21719. By applying equation (1) Message is generated by KGC, and the generated messages are M=55131157819910041417986790291061886647009280000 along with the group key K=22971 is manipulated according to equation (1) and then it is broadcasted to all the ten members in the group.

After receiving this message, member1 computes the key (K) by equating (2) as M % (P1 ⊕ S1). i.e.,K=55131157819910041417986790291061886647009302971 % (55387 ⊕ 28931). The value obtained from the above calculated results in the original key value which is 22971. Similarly all the other members in the group calculate their keys by doing M % (Pi ⊕ Si), i=2, 3...10. We have taken K (Key) sizes as 64,128, 512, 1024 bits and the value of S (prime) has been taken has 64, 128, 512 and 1024 bits. When a non-group member mk attempts to compute the group key with a value $P_k$=43651, $S_k$= 45079 and the above message M. The member $m_k$ will not able to calculate the key. i.e., M % ($P_k$ ⊕ $S_k$) = 51311578199100414179867902910618886647009302971 % (43651 ⊕ 45079) = 2559≠22971≠K.

## 5. PERFORMANCE ANALYSIS

The performance analysis of the proposed and implemented system is measured with respect to the size of group members. As the group size increases the value of Message generation time and Key Extraction time gets increased. The observed values are plotted in a graph. The performance of the algorithm for generating the message M for different group size is shown in Fig.3. From the graph shown it is observed that the proposed technique is more efficient and takes less time for generating message for small groups.

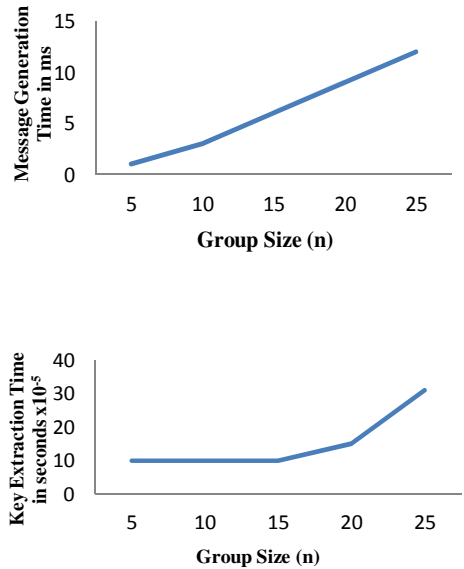

Figure. 3 (a) Message generation for small groups (b) Key extraction for small groups

The Message Generation time is measured in milliseconds (ms). Moreover, it is found that the efficiency of the algorithm for extracting the key takes less time. Also, it is observed that when the group size increases the Key Extraction time also increases rapidly. The strength of the RSA algorithm relies on the fact that the given M (product of two large prime numbers), it is not possible to find the two factors in polynomial time (It is NP-hard). In this proposed algorithm, it has more complex message M, than what is used in RSA algorithm. Therefore, it is very difficult to find any of the prime numbers $P_1$, $P_2$,.., $P_n$ is NP-hard. Thus the Key K is more secure and safe preventing it from man-in-the-middle attack and brute-force attack. To maintain the forward and backward secrecy, KGC generates a new message and broadcast it everytime when a member join or leaves the system.

## 5. CONCLUSION

Key transfer protocol relies on a trusted Key Generation Center (KGC) to select group key and to distribute group keys to all group entities in a secret manner. KGC assign a large prime number to each member in the group. Since the chosen message M as more complex which is compared with the RSA algorithm, the proposed technique is proved to be more efficient key transfer protocol used for group communication. Our proposed algorithm is efficient both in terms of message generation and key extraction. In future, attempts can be made to implement the design for communication with dynamic and hierarchical groups.

**Authors**

R.Velumadhava Rao received the B.E degree in Computer Science and Engineering from C.I.E.T, Coimbatore and M.E degree in Software Engineering from College of Engineering, Guindy, Anna University, Chennai, India. He is currently working as an Assistant Professor in Rajalakshmi Institute of Technology, Chennai.

K.Selvamani received the B.E degree in Electrical and Electronics Engineering from Annamalai University, Chidambaram and M.E degree in Computer Science and Engineering from Bharathiyar University, Coimbatore in December 2000 and a Ph.D degree in Computer Science and Engineering from Anna University in 2011. He is currently working as a Assistant Professor in College of Engineering, Guindy, Anna University, Chennai. His research interest include Web Security, Cryptography and Network Security.

R. Elakkiya received her B.E degree in Computer Science and Engineering from Periyar Maniammai college of technology for women, Thanjavur and M.E degree in Software Engineering from College of Engineering Guindy, Ana University, Chennai, India. She is currently working as an Assistant Profesor in Jerusalem College of Engineering, Chennai.